\documentclass[twocolumn, twoside, conference, 10pt]{IEEEtran}
\usepackage{amsmath,amssymb,graphicx,cite,Comment,psfrag}

\begin{document}

\setcounter{page}{0}

\title{Low-Complexity LDPC Codes with Near-Optimum Performance over the BEC}
\author{\IEEEauthorblockN{Enrico Paolini, Michela Varrella and Marco Chiani}
\IEEEauthorblockA{DEIS, WiLAB\\
University of Bologna\\
via Venezia 52, 47023 Cesena (FC), Italy\\
\footnotesize{\texttt{\{e.paolini,michela.varrella,marco.chiani\}@unibo.it}}}
\and
\IEEEauthorblockN{Balazs Matuz and Gianluigi Liva}
\IEEEauthorblockA{Institute of Communication and Navigation\\
Deutsches Zentrum fÄur Luft- und Raumfahrt (DLR)\\
82234 Wessling, Germany\\
\footnotesize{\texttt{\{Balazs.Matuz,Gianluigi.Liva\}@dlr.de}}}}
\date{\today}
\thispagestyle{empty} \setcounter{page}{1} \maketitle
%
%
\begin{abstract}
Recent works showed how low-density parity-check (LDPC) erasure
correcting codes, under maximum likelihood (ML) decoding, are
capable of tightly approaching the performance of an ideal
maximum-distance-separable code on the binary erasure channel. Such
result is achievable down to low error rates, even for small and
moderate block sizes, while keeping the decoding complexity low,
thanks to a class of decoding algorithms which exploits the
sparseness of the parity-check matrix to reduce the complexity of
Gaussian elimination (GE). In this paper the main concepts
underlying ML decoding of LDPC codes are recalled. A performance
analysis among various LDPC code classes is then carried out,
including a comparison with fixed-rate Raptor codes. The results
show that LDPC and Raptor codes provide almost identical performance
in terms of decoding failure probability vs. overhead.
\end{abstract}

\section{Introduction}\label{sec:INTRO}
\PARstart{L}{ow}-density parity-check codes
\cite{studio3:GallagerBook} exhibit extraordinary performance under
iterative (IT) decoding over a wide range of communication channels.
It was even proved that some classes of LDPC code ensembles can
asymptotically approach, under IT decoding, the binary erasure
channel (BEC) capacity with an arbitrarily small gap
\cite{pfister2005:bounded}. However, problems arise when using these
asymptotically optimal constructions in conjunction with a
finite-length $(n,k)$ LDPC code. In fact, the corresponding IT
performance curve though quite good at high error rates, denotes a
coding gain loss with respect to that of an ideal code matching the
Singleton bound. Furthermore, at low error rates the performance
curve deviates even more from the ideal behavior due to the error
floor phenomenon caused, for IT decoding, by small size stopping
sets. In general, lowering the IT error floor implies a sacrifice in
terms of coding gain respect to the Singleton bound at high error
rates.

It is well known that failures of the IT decoder over the BEC are
due to stopping sets. Since there exist sets of variable nodes (VNs)
representing stopping sets for the IT decoder, but not for the ML
decoder, a decoding strategy consists in performing IT decoding and,
upon an IT decoder failure, employing the ML decoder to try to
resolve the residual stopping set. This hybrid decoder achieves the
same performance as ML. The performance curve obtained after the ML
step approaches the Singleton bound curve more closely than that
relative to IT decoding and down to lower error rates. In fact, the
error floor under ML decoding only depends on the distance spectrum.

If the communication channel is a binary erasure channel (BEC), ML
decoding is equivalent to solving the linear equation
\begin{align}\label{eq:ML-decoding}
\mathbf{x}_{\overline{K}} \, \mathbf{H}^{T}_{\overline{K}} =
\mathbf{x}_K \, \mathbf{H}^{T}_K,
\end{align}
where $\mathbf{x}_{\overline{K}}$ ($\mathbf{x}_K$) denotes the set
of erased (correctly received) encoded bits and
$\mathbf{H}_{\overline{K}}$ ($\mathbf{H}_K$) the submatrix composed
of the corresponding columns of the parity-check matrix
$\mathbf{H}$. Then, ML decoding for the BEC can be implemented as a
Gaussian elimination performed on the binary matrix
$\mathbf{H}^{T}_{\overline{K}}$ : its complexity is in general
$O(n^3)$, where $n$ is the codeword length. It is obvious that for
long block lengths ML decoding becomes impractical so that IT
decoding is preferred. For LDPC codes, it is indeed possible to take
advantage of both the ML and IT approach. To keep complexity low, a
first decoding attempt is done in an iterative manner
\cite{studio3:MLcommlet}. If not successful, the residual set of
unknowns is processed by an ML decoder. Efficient ways of
implementing ML decoders for LDPC codes can be found in
\cite{miller04:bec}, whose approach takes basically benefit from the
sparse nature of the parity-check matrix of the code.

A thorough performance analysis of LDPC codes under
reduced-complexity ML decoding is provided in this paper. We use a
class of fixed-rate Raptor codes as a benchmark for our performance
evaluations. These fixed-rate Raptor codes are obtained from the
rate-less codes recommended for the Multimedia Broadcast Multicast
Service (MBMS) by selecting \emph{a priori} the codeword length $n$.
Raptor codes are universally recognized as the state-of-the-art
codes for the BEC, and they are currently under investigation for
fixed-rate applications within the Digital Video Broadcasting (DVB)
standards family \cite{DVB-SH}. As for LDPC codes, also for Raptor
codes efficient ML decoders are available \cite{MBMS05:raptor}.

The outcomes presented in this paper are of great interest for many
different applications, such as those listed next.

\begin{itemize}
\item \emph{Wireless video/audio streaming.} Link-layer coding is currently
applied to the video streams in the framework of the DVB-H/SH
standards. In such a context, erasure correcting codes take care of
the fading mitigation, which is crucial especially in the case of
mobile users, in challenging propagation environments
(urban/suburban and land-mobile-satellite channels).
Capacity-approaching performance is here highly desired in order to
increase the service availability. Mobile applications require
low-complexity decoders as well.

\item
\emph{File delivery in broadcasting/multicasting networks.} Reliable
file delivery in broadcasting/multicasting networks finds a
near-optimal solution in erasure correcting codes. In such a
scenario, reliability cannot be guaranteed by any automatic repeat
request (ARQ) mechanism, due to the broadcast nature of the channel.
Erasure correcting codes would limit (or avoid) the usage of packet
retransmissions.

\item  \emph{File delivery in
point-to-point communications.} Also in point-to-point links, file
delivery may require further protection at upper layers. This is
true especially if retransmissions are impossible (due to the
absence of a return channel or due to long round-trip delays).

\item
\emph{Deep space communications.} Deep space communication has been
always an ideal application field for error correcting codes. The
Consultative Committee for Space Data Systems (CCSDS) is currently
investigating the adoption of erasure correcting codes to further
protect the telemetry down-link, especially for deep-space missions,
which are not suitable for ARQ. In such a context, the possibility
of processing the data off-line, together with the relatively-low
data rates (up to some Mbps), makes ML decoding of linear block
codes a concrete solution, even in absence of low-complexity
decoders. A mandatory feature is instead represented by
low-complexity encoder implementations.
\end{itemize}

The paper is organized as follows. In Section \ref{sec:LDPCML}
reduced-complexity ML decoding of LDPC codes on the binary erasure
channel is reviewed, including some insights on the code design for
ML. In Section \ref{sec:raptor} a class of fixed-rate Raptor codes
is introduced, together with a summary on their ML encoder/decoder
implementations. Section \ref{sec:RESULTS} provides simulation
results for both LDPC and fixed-rate Raptor codes. Conclusions
follow in Section \ref{sec:CONCLUSIONS}.

\section{LDPC codes and ML decoding}\label{sec:LDPCML}

This section is organized in a two-fold way. First, the main
concepts of the ML decoder of \cite{miller04:bec} will be explained.
Second, new code designs for ML will be presented and evaluated with
regard to complexity and performance.

\subsection{Efficient maximum-likelihood decoding for LDPC codes over the erasure channel}\label{subsec:ML-decoder}
The problem of GE over large sparse, binary matrices has been widely
investigated in the past. A common approach relies on structured GGE
with the purpose of converting the given system of sparse linear
equations to a new smaller system that can be solved afterwards by
brute-force GE\cite{odlyzko84discrete,studio3:RiUr01,miller04:bec}.
Here, we'll provide an simplified overview of the approach presented
in \cite{miller04:bec}.
For sake of clarity, let's apply column permutations to arrange the
parity check matrix $\mathbf{H}$ as in \eqref{eq:ML-decoding}: the
left part shall contain all the columns related to known variable
nodes ($\mathbf{H}_{K}$), whereas the right part shall be made up of
all the columns related to erased variable nodes
($\mathbf{H}_{\overline{K}}$). Thus, to solve the unknowns, we
proceed as follows:

\begin{figure*}[]
\begin{center}
\includegraphics[width=11.4cm,draft=false]{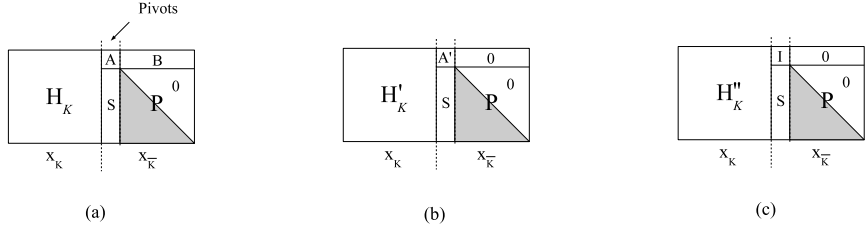}
\end{center}
\caption{ML decoder as in \cite{miller04:bec}. (a) Pivots selection
within $\mathbf{H}_{\overline{K}}$. (b) Zeroing of matrix
$\mathbf{B}$. (c) Gaussian Elimination on
$\mathbf{A}'$.}\label{fig:matrices}
\end{figure*}

\begin{itemize}
\item Perform diagonal extension steps on $\mathbf{H}_{\overline{K}}$.
This results in the sub-matrices $\mathbf{B}$, as well as
$\mathbf{P}$ that is in a lower triangular form, and columns that
cannot be put in lower triangular form (columns of matrices
$\mathbf{A}$ and $\mathbf{S}$). The variable nodes corresponding to
the former set of columns build up the so-called pivots (see Figure
\ref{fig:matrices}(b)). Note that all remaining unknown variable
can be obtained by linear combination of the pivots and of the known
variables.
\item Zero the matrix $\mathbf{B}$ which elements can be expressed by the sum
of the pivots (Figure \ref{fig:matrices}(a)).
\item Resolve the system by performing Gaussian elimination only on $\mathbf{A}'$.
Out of the pivots the unknown variables can be obtained quite easily
due to the lower triangular structure of $\mathbf{P}$.
\end{itemize}

It should be obvious that the main strength of this algorithm lies
in the fact that GE is only performed on $\mathbf{A}'$ and not on
the entire set of unknown variables. Therefore, it is of great
interest to keep the dimensions of $\mathbf{A}'$ rather small. This
can be obtained by sophisticated ways of choosing the pivots
\cite{miller04:bec} and by a judicious code design
\cite{studio3:MLcommlet}. Besides, to reduce the complexity further
the brute-force Gaussian elimination step on $\mathbf{A}'$ could be
replaced by other algorithms.

Note that the ML decoder for an $(n,k)$ LDPC code operates on a
sparse matrix with at most $n-k$ columns and $n-k$ rows. The
relevance of this consideration will become more clear after the
description of the ML Raptor decoder \cite{MBMS05:raptor} provided
in Section \ref{sec:raptor}.

\subsection{On the code design}
\begin{figure}[h]
\begin{center}
\includegraphics[width=\columnwidth,draft=false]{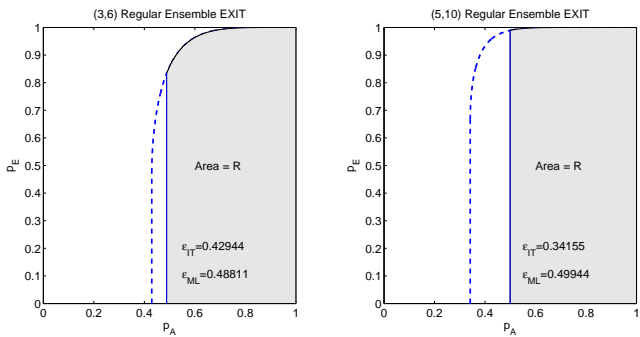}
\end{center}
\caption{EXIT functions for the (3,6) and the (5,10) regular LDPC
ensembles. Dashed lines represent the iterative decoder EXIT
function. Solid lines are placed in correspondence of the ML
thresholds upper bounds.}\label{fig:GEXIT}
\end{figure}

The usual code design employed for LDPC codes over BEC deals with
the selection of proper degree distributions (or protographs)
achieving high iterative decoding thresholds $\epsilon_{IT}$ (as
close as possible to the limit given by $1-R$). A $(n,k)$ LDPC code
is then picked from the ensemble defined by the above-mentioned
degree distributions. The selection may be performed following some
girth optimization techniques. Such an iterative-decoding-based
design criterion does not answer to the need of finding good codes
for ML decoding. Namely, a different figure shall be put in the
focus of the degree distribution optimization. In our code design,
the corresponding feature of $\epsilon_{IT}$  under ML decoding,
i.e., the ML decoding threshold $\epsilon_{ML}$, is the subject of
the figure driving the optimization. A method for deriving a tight
upper bound on the ML threshold for an LDPC ensemble can be found in
\cite{Montanari:lifeabovethreshold}. The upper bound on
$\epsilon_{ML}$ can be derived as follows.
\begin{itemize}
\item Consider an $(n,k)$ LDPC code and its corresponding
IT decoder. The extrinsic information transfer (EXIT) curve of the
code (under IT decoding) can be derived in terms of \emph{extrinsic
erasure probability} at the output of the decoder ($p_E$) as a
function of the \emph{a priori erasure probability} (input of the
decoder, $p_A$). For $n\rightarrow +\infty$, the EXIT curve of the
ensemble defined by $\lambda(x)$ and $\rho(x)$ is a function of the
degree distributions, and can be obtained in parametric form as
\begin{equation}
p_A=\frac{x}{\lambda(1-\rho(1-x))}\label{eq:EXIT1}
\end{equation}
\begin{equation}
p_E=\Lambda(1-\rho(1-x))\label{eq:EXIT2}
\end{equation}
with $x\in [x_{BP}, 1]$, being $x_{BP}$ the value of $x$ for which
$p_A=\epsilon_{BP}$, and $\Lambda(x)=\sum\Lambda_i x^i$, being
$\Lambda_i$ the fraction of variable nodes with degree $i$. EXIT
functions of regular LDPC ensembles are displayed in Figure
\ref{fig:GEXIT} (dashed lines).
\item Due to the Area Theorem \cite{Ashikhmin:AreaTheorem}, the area
below the EXIT function of the code, under ML decoding, must equal
the code rate $R$. Note that the EXIT function defined by
\eqref{eq:EXIT1},\eqref{eq:EXIT2} is IT-decoder-based. Hence, the
are below the EXIT curve might be larger than the code rate.
\item Consider the extrinsic erasure probability at the output of an
ML and of an IT decoder. Obviously, $p_E^{ML}\leq p_E^{IT}$.
\item Therefore, by drawing a vertical line on the EXIT
function plot of the ensemble, in correspondence with $p_A=p_A^*$,
such that
\[
\int_{p_A^*}^1 p_E(p_A)dp_A=R,
\]
we obtain an upper bound on the ML threshold, i.e.,
$\epsilon_{ML}\leq p_A^*$. For regular LDPC ensembles, see the
example in Figure \ref{fig:GEXIT}.
\end{itemize}

\begin{figure}[h]
\begin{center}
\includegraphics[width=0.7\columnwidth,draft=false]{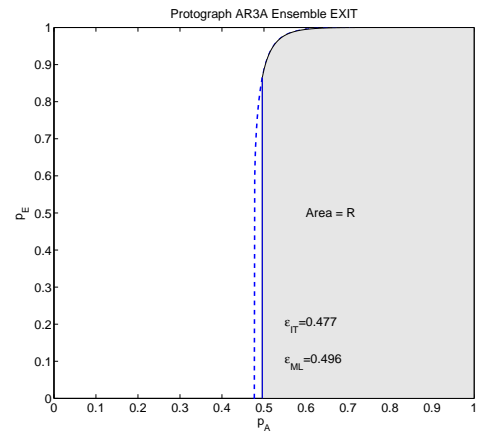}
\end{center}
\caption{EXIT function for the  accumulate-repeat-accumulate
ensemble. Dashed lines represent the iterative decoder EXIT
function. Solid lines are placed in correspondence of the ML
thresholds upper bounds.}\label{fig:GEXIT_ARA}
\end{figure}

In \cite{Montanari:lifeabovethreshold} it was shown that this bound
is extremely tight for regular LDPC ensembles, and for ensembles
whose IT EXIT curve presents one jump (for further details, see
\cite{Montanari:lifeabovethreshold}). Slightly different (but still
rather simple) techniques to obtain tight bounds are applicable also
in the other cases \cite{Montanari:lifeabovethreshold}. Extensions
to the above-mentioned techniques can be applied to other code
ensembles, once the IT EXIT curve is provided. For protograph LDPC
ensembles \cite{studio3:proto_LDPC}, a rather simple approach would
then be the application of the protograph EXIT analysis of
\cite{studio3:proto_exit2007} to obtain the IT EXIT curve for a
given protograph ensemble. The upper bound on the maximum-likelihood
threshold can then be obtained as for conventional $(\lambda,\rho)$
ensembles. An example of the IT EXIT curve for an
accumulate-repeat-accumulate (ARA) ensemble
\cite{studio3:ARAGlobecomm} is provided in Figure
\ref{fig:GEXIT_ARA}, as well as the derivation of the related ML
threshold upper bound. Proofs on the tightness of the bound for
protograph ensembles are currently missing and are not in the scope
of this paper.

For the regular ensembles, the improvement given by the ML decoder
is usually large (see Table \ref{Tab:Reg_Thresholds}).

A rule of thumb for the design of capacity-approaching LDPC codes
under ML consists in the selection of sufficiently dense
parity-check matrices, by keeping for instance a relatively large
average check node degree. To given an idea, we found that for rate
$1/2$ LDPC ensembles, an average check node degree $d_c\geq 9$ is
sufficient to provide ML thresholds close to the Shannon limit
\cite{studio3:MLcommlet}. This heuristic rule seems to work for both
regular and irregular ensembles.

\begin{table}[]
\begin{center}
\caption{ML and IT decoding threshold for regular LDPC ensembles vs.
the Shannon limit, $\epsilon_{Sh}$. }
\begin{tabular}{|c|c|c|c|}
  \hline
  Ensemble & $\epsilon_{ML}$ & $\epsilon_{IT}$ & $\epsilon_{Sh}$ \\
  \hline
  (3,6) & $0.4881$ & $0.4294$ & $0.5000$ \\
  (4,8) & $0.4977$ & $0.3834$ & $0.5000$ \\
  (5,10) & $0.4994$ & $0.3416$ & $0.5000$ \\
  (6,12) & $0.4999$ & $0.3075$ & $0.5000$ \\
  (3,9) & $0.3196$ & $0.2828$ & $0.3333$ \\
  (4,12) & $0.3302$ & $0.2571$ & $0.3333$ \\
  (5,15) & $0.3324$ & $0.2303$ & $0.3333$ \\
  \hline
\end{tabular}
\label{Tab:Reg_Thresholds} \end{center}
\end{table}

\subsection{GeIRA codes with low-complexity ML decoder}

In \cite{studio3:MLcommlet}, it is shown that good iterative
decoding thresholds are indeed highly desirable for ML decoding,
since they allows reducing the decoder complexity. More
specifically, in \cite{studio3:MLcommlet} some simple design rules
are provided, leading to codes with good IT thresholds,
near-Shannon-limit ML thresholds, low error floors, with simple
(turbo-code-like) encoders \cite{liva05:CL}. The proposed code
design leads to a class of generalized irregular repeat accumulate
(GeIRA) codes tailor-made for efficient ML decoding. In Section
\ref{sec:RESULTS}, numerical results on GeIRA codes with different
coding rates/block lengths will be provided.

\section{Fixed-rate Raptor codes} \label{sec:raptor}

Raptor codes were introduced by Shokrollahi in
\cite{shokrollahi06:raptor}. They are an instance of the concept of
\emph{fountain code}\footnote{Commonly, the expression ``fountain
code'' is used to refer to a code which can produce on-the-fly any
desired number of encoded symbols from $k$ information symbols.}
\cite{byers02:fountain} and, thanks to the large degrees of freedom
in parameter choice, they can be applied to several systems,
increasing their reliability. Recently, a fully specified version of
Raptor codes has been approved as a means to efficiently disseminate
data over a broadcast network \cite[Annex B]{MBMS05:raptor}. A
$(n,k)$ fixed-rate Raptor code can be obtained by limiting to $n$
the amount of symbols produced by the Raptor encoder. Fixed-rate
Raptor codes derived from the MBMS standard \cite[Annex
B]{MBMS05:raptor} are currently under investigation for the multi
protocol encapsulation (MPE) level protection within the DVB
standards family \cite{DVB-SH}. In the following, we will provide
first a description of the Raptor codes specified in \cite[Annex
B]{MBMS05:raptor}, including some insights on their encoding and
decoding algorithms.

\begin{figure}[h]
  \centering
  \includegraphics[width=0.499\textwidth]{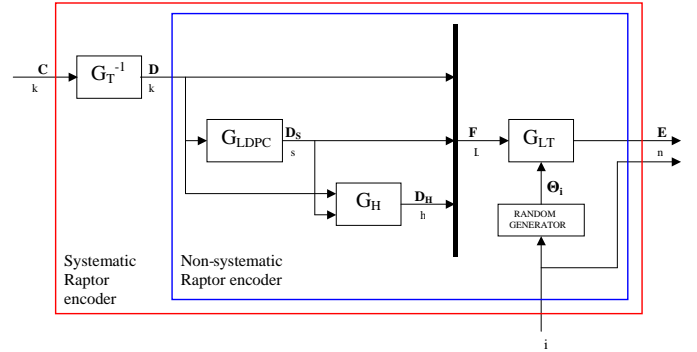}
  \caption{Block diagram of the systematic Raptor encoder specified in \cite{MBMS05:raptor}.}
  \label{Fig:blocchiRaptor}
\end{figure}

The Raptor code can be viewed as the concatenation of several codes.
For example  the Raptor encoder specified in \cite{MBMS05:raptor} is
depicted in Fig. \ref{Fig:blocchiRaptor}. The most-inner code is a
non systematic Luby-transform (LT) code \cite{luby02:LT} with $L$
input symbols $\mathbf{F}$, producing the encoded symbols
$\mathbf{E}$. The symbols $\mathbf{F}$ are known as
\emph{intermediate symbols}, and are generated through a pre-coding,
made up of some outer high-rate block coding, effected on the $k$
symbols $\mathbf{D}$. The $s$ intermediate symbols $\mathbf{D}_s$
are known as \emph{LDPC symbols}, while the $h$ intermediate symbols
$\mathbf{D}_h$ are known as \emph{half symbols}. The combination of
pre-code and LT code produces a non systematic Raptor code. The
parameters $s$ and $h$ are functions of $k$, according to
\cite{MBMS05:raptor}. Some pre-processing is to be put before the
non-systematic Raptor encoding to obtain a systematic one. Such a
pre-processing consists in a rate-1 linear code generating the $k$
symbols $\mathbf{D}$ from the $k$ information symbols $\mathbf{C}$.

LT codes are the first practical implementation of fountain codes.
An unique encoded symbol ID (ESI) is assigned to each encoded
symbol. Starting from an ESI $i$, the encoded symbol $E_{i}$ is
computed by xor-ing a subset $\mathbf{\Theta}_{i}$ of $d_i$
intermediate symbols. The number $d_i$, known as the \emph{degree}
associated with the encoded symbol $E_{i}$, is a random integer
between $1$ and $L$: the $d_i$ intermediate symbols are chosen at
random according to a specific probability distribution. As a
consequence, in order to recover the information symbols the decoder
needs both the set of encoded symbols $E_{i}$ and of the
corresponding $\mathbf{\Theta}_{i}$. This last information can
either be explicitly transmitted or obtained by the decoder through
the same pseudo-random generator used for the encoding, starting
from ESIs, which have therefore to be sent together with the
corresponding encoded symbols.

Some of the main properties of LT codes are that the encoder can
generate as many encoded symbols as desired and that the decoder is
able to recover the block of source symbols from any set of received
encoded symbols, whose number is only slightly greater than that of
the source symbols (in fact the code claims a low level of
overhead). A Raptor code, whose core consists of an LT code, inherit
such properties. The addiction of a pre-coding phase is used to
obtain an encoding/decoding complexity linear with $k$; a feature
which is missing in the mere LT code.

\subsection{Fixed-rate Raptor generator matrix}\label{subsec:raptor-G}

Considering a systematic Raptor code as a finite length $(n,k)$
linear block code (fixed-rate Raptor code), we can ask what is the
structure of its generator matrix. This problem is addressed next
for the Raptor code specified in
\cite{MBMS05:raptor}\footnote{Throughout this section, the vectors
are intended as column vectors (unless explicitly mentioned) and the
generator matrix of a $(n,k)$ linear block code is expressed as a
$(n \times k)$ matrix.}.
The generator matrix of the first pre-coding stage is given by
$[\mathbf{I}_k | G_{\textrm{LDPC}}^T]^T$. According to the
specifications in \cite{MBMS05:raptor}, $\mathbf{G}_{\textrm{LDPC}}$
consists in columns all of weight equal to $3$, regardless the value
of $k$. On the other hand, the generator matrix of the second
pre-coding stage is given by $[\mathbf{I}_{s+k} |
\mathbf{G}_H^T]^T$, where $\mathbf{G}_{\textrm{H}}$ is a $(h \times
(s+k))$ matrix consisting in columns all of constant weight: each
column is an element of the Grey sequence of weight $h'$, where $h'
= \lceil h/2 \rceil$.
Finally, let us denote by $\mathbf{G}_{\textrm{LT}}$ the $(n \times
L)$ LT code generator matrix (regarded as a finite length $n$ linear
block code). It is built in such a way that the row of index $i$ has
$d_{i}$ ones in $\mathbf{\Theta}_{i}$ positions, where $d_{i}$ and
$\mathbf{\Theta}_{i}$ are derived from the ESI $i$, through
pseudo-random algorithms described in in \cite{MBMS05:raptor}. Next,
we use the notation
$\mathbf{G}_{\textrm{LT}}(i_{1},i_{2},\dots,i_{r})$ to denote the
$(r \times L)$ submatrix of $\mathbf{G}_{\textrm{LT}}$ composed of
the rows with indexes $(i_{1},i_{2},\dots,i_{r})$. The notation
$\mathbf{G}_{\textrm{LT}}$ is equivalent to
$\mathbf{G}_{\textrm{LT}}(1,\dots,n)$.

The $L=k+s+h$ intermediate symbols $\mathbf{F}$ are obtained from
$\mathbf{D}$ as
\begin{equation*}
    \mathbf{F}= \left[ \begin{array}{l}
    \mathbf{D} \\
    \mathbf{D}_{s} \\
    \mathbf{D}_{h}
    \end{array}  \right]
\end{equation*}
\noindent through the relations
\begin{align}\label{Eq:pre-coding-1}
  \mathbf{D}_{s} = \mathbf{G}_{\textrm{LDPC}}\cdot \mathbf{D}
\end{align}
\begin{align}\label{Eq:pre-coding-2}
  \mathbf{D}_{h} = \mathbf{G}_{\textrm{H}} \cdot \left[ \begin{array}{l}
                                                \mathbf{D} \\
                                                \mathbf{D}_{s}
                                                \end{array}
                                                \right]\, .
\end{align}

\noindent 

The intermediate symbols $\mathbf{F}$ are the inputs to the LT
encoder for deriving the $n$ encoded symbols $\mathbf{E}$ as
\begin{equation}
    \mathbf{E}  =  \mathbf{G}_{\textrm{LT}}\cdot \mathbf{F} .
    \label{Eq:LTencoding}
\end{equation}

\noindent Let us subdivide $\mathbf{G}_{\textrm{LT}}$ as

    \[
    \mathbf{G}_{\textrm{LT}}= \left[ \begin{array}{c|c|c}
                           \mathbf{G}^{\textrm{I}}_{\textrm{LT}} & \mathbf{G}^{\textrm{II}}_{\textrm{LT}} & \mathbf{G}^{\textrm{III}}_{\textrm{LT}} \\
                           \end{array}  \right]\, ,
\]

\noindent where the sizes of the three submatrices are $(n \times
k)$, $(n \times s)$ and $(n \times h)$, respectively.
If also $\mathbf{G}_{\textrm{H}}$ is subdivided as

\[
    \mathbf{G}_{\textrm{H}}= \left[ \begin{array}{c|c}
                           \mathbf{G}^{\textrm{I}}_{\textrm{H}} & \mathbf{G}^{\textrm{II}}_{\textrm{H}}  \\
                           \end{array}  \right]\, ,
\]
that is into two submatrices whose sizes are $(h \times k)$ and $(h
\times s)$, respectively, then the non-systematic Raptor code
generator matrix can be expressed as

\begin{align*}
    \mathbf{G}_{\textrm{R,n-sys}} & = \mathbf{G}^{\textrm{I}}_{\textrm{LT}} + \mathbf{G}^{\textrm{II}}_{\textrm{LT}} \cdot \mathbf{G}_{\textrm{LDPC}} \notag \\
    \, & \qquad + \mathbf{G}^{\textrm{III}}_{\textrm{LT}} \left(  \mathbf{G}^{\textrm{I}}_{\textrm{H}} + \mathbf{G}^{\textrm{II}}_{\textrm{H}}\cdot \mathbf{G}_{\textrm{LDPC}} \right)
\end{align*}
which satisfies the relation:

\begin{equation}
    \mathbf{E}  =  \mathbf{G}_{\textrm{R,n-sys}}\cdot \mathbf{D} . \nonumber
\end{equation}

Let's now subdivide $\mathbf{G}_{\textrm{R,n-sys}}$ into the two
submatrices $\mathbf{G}^{\textrm{I}}_{\textrm{R,n-sys}}$ and
$\mathbf{G}^{\textrm{II}}_{\textrm{R,n-sys}}$, whose sizes are $(k
\times k)$ and $\left((n-k) \times k \right)$, respectively:
\[
    \mathbf{G}_{\textrm{R,n-sys}}= \left[ \begin{array}{c}
                           \mathbf{G}^{\textrm{I}}_{\textrm{R,n-sys}} \\
                            \mathbf{G}^{\textrm{II}}_{\textrm{R,n-sys}}  \\
                            \end{array}  \right] .
\]
For a systematic code it must be valid the following
\[
     \begin{array}{lr}
       E_{i} \equiv C_{i} \phantom{.} & \phantom{.} \forall
       i=1,...,k \, ,
     \end{array}
\]
and therefore

\begin{align}\label{eq:E}
    \left[ \begin{array}{l}
        \mathbf{G}^{\textrm{I}}_{\textrm{R,n-sys}} \\
        \mathbf{G}^{\textrm{II}}_{\textrm{R,n-sys}} \\
    \end{array}  \right]  \cdot \mathbf{D} & = \left[ \begin{array}{l}
                                                 \mathbf{E}_{[1,..,k]} \\
                                                 \mathbf{E}_{[k+1,..,n]}
                                                \end{array}  \right] \\
                                                \, & = \left[ \begin{array}{c}
                                        \mathbf{C} \\
                                        \mathbf{E}_{[k+1,..,n]}
                                        \end{array}  \right]  .
\end{align}

\noindent We have introduced in \eqref{eq:E} the notations
$\mathbf{E}_{[1,..,k]}$ and $\mathbf{E}_{[k+1,..,n]}$ to denote the
first $k$ and the last $n-k$ encoded symbols, respectively.

We can state that the pre-processing matrix generating $\mathbf{D}$
from $\mathbf{C}$ can be obtained by
\begin{equation*}
    \mathbf{G}_{\textrm{T}}^{-1}=(\mathbf{G}^{\textrm{I}}_{\textrm{R,n-sys}})^{-1}
\end{equation*}
and, as a consequence, the systematic Raptor code generator matrix
is
\begin{equation}
    \mathbf{G}_{\textrm{R,sys}}= \left[ \begin{array}{c}
        \mathbf{I}_{k} \\
        \mathbf{G}^{\textrm{II}}_{\textrm{R,n-sys}}
    \end{array}  \right]
    \label{Eq:RaptorGeneratorMatrix}
\end{equation}
In \eqref{Eq:RaptorGeneratorMatrix} $\mathbf{I}_{k}$ denotes the $(k
\times k)$ identity matrix. Obviously,
$\mathbf{G}^{\textrm{I}}_{\textrm{R,n-sys}}$ can be inverted if and
only if it has full rank $k$. By initializing the random generator
of inner LT code through the so-called \emph{systematic index}
(defined in \cite{MBMS05:raptor}), this property is fulfilled for
all $k=4,\dots,8192$.

\subsection{Raptor Encoding}\label{subsec:Raptor-encoding}

%
%
%
\begin{figure}[t]
  \centering
  \psfrag{s}[cb]{$s$} \psfrag{k}[cb]{$k$}  \psfrag{h}[cb]{$h$} \psfrag{n}[cb]{$n$}
  \psfrag{1}[cb]{$\mathbf{G}_{\textrm{LDPC}}$}
  \psfrag{2}[lb]{$\mathbf{I}_s$} \psfrag{3}[lb]{$\mathbf{Z}$}
  \psfrag{4}[cb]{$\mathbf{G}_{\textrm{H}}$}
  \psfrag{5}[lb]{$\mathbf{I}_h$} \psfrag{6}[cb]{$\mathbf{G}_{\textrm{LT}}(1,\dots,n)$}
  \includegraphics[width=0.25\textwidth,angle=270]{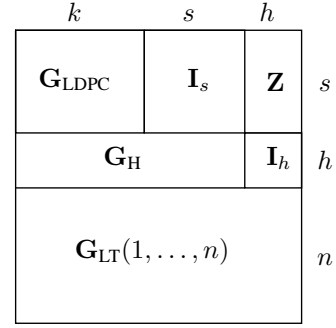}
  \caption{Structure of the encoding matrix $\mathbf{A}$
  for an $(n,k)$ Raptor code specified in \cite{MBMS05:raptor} ($L=k+s+h$).}
  \label{Fig:A}
\end{figure}

The relations \eqref{Eq:pre-coding-1}, \eqref{Eq:pre-coding-2} and
\eqref{Eq:LTencoding} can conveniently be represented as:

    \[
    \mathbf{A}\cdot \mathbf{F} = \left[ \begin{array}{c}
                                                 \mathbf{0} \\
                                                 \mathbf{E}_{[1,..,n]}
                                                \end{array}  \right]
\]
whereby $\mathbf{A}$ is a $((s+h+n) \times (s+h+k))$ binary matrix
called \emph{encoding matrix}, whose structure is shown in Fig.
\ref{Fig:A}. In this figure, $\mathbf{I}_{s}$ is the $(s \times s)$
identity matrix, $\mathbf{I}_{h}$ is the $(h \times h)$ identity
matrix and $\mathbf{Z}$ is the $(s \times h)$ all-zero matrix. The
matrix $\mathbf{A}$ doesn't properly represent the Raptor code
generator matrix (which is defined in
\eqref{Eq:RaptorGeneratorMatrix} instead), but includes the set of
constraints imposed by the pre-coding and LT coding together. We use
next the notation $\mathbf{A}(i_{1},i_{2},..,i_{r})$ to indicate the
$((s+h+r) \times L)$ submatrix of $\mathbf{A}$ obtained by selecting
only the rows of $\mathbf{G}_{\textrm{LT}}$ with indexes
$(i_{1},i_{2},..,i_{r})$. Again, $\mathbf{A}$ is equivalent to
$\mathbf{A}(1,\dots,n)$.

A possible Raptor encoding algorithm exploits a submatrix of
$\mathbf{A}$. %
%
Such a matrix, consisting of the first $L$ rows of $\mathbf{A}$, is
used to obtain $\mathbf{F}$ solving the system of linear equations:

\begin{equation*}
    \mathbf{A}(1,...,k)\cdot \mathbf{F} = \left[ \begin{array}{c}
                                                 \mathbf{0} \\
                                                 \mathbf{\mathbf{C}}
                                                \end{array}  \right] .
\end{equation*}

\noindent At this point it is sufficient to multiply $\mathbf{F}$ by
the LT generator matrix to produce the encoded symbols $\mathbf{E}$,
according to \eqref{Eq:LTencoding}.

\subsection{Raptor Decoding}\label{subsec:Raptor-decoding}

The most direct way to decode the received sequence lies in
inverting each encoding step of Fig.~\ref{Fig:blocchiRaptor}; in
this case you work on individual sub-codes. When using ML decoding
at each sub-code, such a method requires the inversion of a matrix
for each code, so it doesn't appear to be the best solution from the
computational viewpoint \cite{Luby2006:Raptor}. Moreover, if the
number of received encoded symbols is not larger enough than (which
in many cases may mean much higher than) the number of source symbol
$k$, it shows an high failure probability.

For example, let's assume that only a subset of encoded symbols of
ESIs $(i_{1},i_{2},...,i_{r})$ are available at the decoder. The
first step the decoder should perform is to solve the system of
linear equations:
\begin{displaymath}
    \mathbf{G}_{\textrm{LT}}(i_{1},i_{2},..,i_{r})\cdot \mathbf{F} = \mathbf{E}_{[i_{1},i_{2},..,i_{r}]} .
\end{displaymath}

The matrix $\mathbf{G}_{\textrm{LT}}(i_{1},i_{2},..,i_{r})$ has $(r
\times L)$ size and, obviously, the necessary condition to solve the
system is that $r \geq L$. If such a condition is not fulfilled, the
decoding fails. It means that to recover the source symbols the
decoder requires at least $L$ encoded symbols (let's recall that
$L=k+s+h$).

Such a method doesn't exploit the fact that the $L$ intermediate
symbols are not independent from each other, but subject to the
pre-coding constraints, instead. Therefore, to obtain the
intermediate symbols $\mathbf{F}$ by using a submatrix of
$\mathbf{A}$ (which consider such constraints) turns out to be a far
more efficient solution.

According to the above-mentioned assumption, the first decoding step
will turn into:
\begin{displaymath}
    \mathbf{A}(i_{1},i_{2},..,i_{r})\cdot \mathbf{F} = \left[ \begin{array}{c}
                                                 \mathbf{0} \\
                                                 \mathbf{E}_{[i_{1},i_{2},..,i_{r}]}
                                                \end{array}  \right]
\end{displaymath}

\noindent where $\mathbf{A}(i_{1},i_{2},..,i_{r})$ is a $((s+h+r)
\times L)$ matrix, as defined above. The system can be solved by
Gaussian elimination (ML decoding) only if $s+h+r \geq L$, that is
$r \geq k$ (note that this is a necessary condition for successful
ML decoding, not a sufficient one). In this way the number of
encoded symbols required at the decoder is definitely lower compared
to that in the previous case and, notably, is close to the number of
source symbols $k$. Once $\mathbf{F}$ is known, the source symbols
$\mathbf{F}$ are easily recovered by

\begin{displaymath}
    \mathbf{C} = \mathbf{G}_{\textrm{LT}}(1,...,k)\cdot \mathbf{F} .
\end{displaymath}

\begin{figure*}[]
  \centering
  \includegraphics[width=0.7\textwidth]{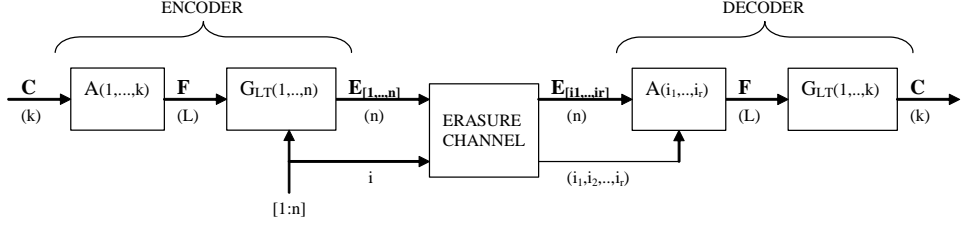}
  \caption{Overview of the encoding and decoding process for the systematic Raptor code specified in \cite{MBMS05:raptor}.}
  \label{Fig:CodDec}
\end{figure*}

\noindent To sum up, when the described encoding and decoding
algorithms are employed, both the encoding and the decoding are
performed by making use of operations which are analogous in the two
case (Fig. \ref{Fig:CodDec}).

\subsection{Some remarks on the decoding complexity of LDPC and
fixed-rate Raptor codes}

If we take into consideration the first decoding step, an algorithm
to perform GE in a more efficient way on
$\mathbf{A}(i_{1},i_{2},..,i_{r})$ has been proposed in \cite[Annex
E]{MBMS05:raptor}. This algorithm share some similarities with that
proposed in \cite{miller04:bec} for LDPC codes. In both cases, the
erased symbols are solved by mean of a structured GE, exploiting the
sparse nature of the equations to reduce the size of the matrix on
which brute-force GE is performed. The targets of the structured GE
are $\mathbf{H}_{\overline{K}}$ for LDPC codes and $\mathbf{A}$ for
Raptor codes. Consider now a $(n,k)$ LDPC code and its fixed-rate
Raptor counterpart. Suppose also an erasure pattern (introduced by
the communication channel) leading to a small overhead $\delta$,
i.e., that the amount of correctly received symbols is $k+\delta$.
On the LDPC code side, the structured GE will be performed on
$\mathbf{H}_{\overline{K}}$ with size $(n-k) \times (n-k-\delta)$.
For the Raptor code, the structured GE will work on $\mathbf{A}$
with size $(k+\delta+s+h) \times (k+s+h)$. Hence, while for the LDPC
code the complexity of the ML decoder is driven by $(n-k)$ (i.e.,
the amount of redundancy, thus by the code rate $R$), for the Raptor
code the complexity depends just on $k$ (i.e., it's code rate
independent). The result is that for high rates($R>1/2$) LDPC codes
have an inherent advantage in complexity. On the other hand, for
lower rates Raptor codes shall be preferable.


\section{Numerical results}\label{sec:RESULTS}
In this section, some numerical results will be provided for LDPC
and fixed-rate Raptor codes under ML over the BEC. The performance
is provided in terms of codeword error rate (CER) vs. the channel
erasure probability $\epsilon$. The section is organized in
subsections. First, some performance bounds for a $(n,k)$ linear
block code over the BEC are reviewed. Then the performance of some
moderate-block-length LDPC codes is provided. The comparison with
fixed-rate Raptor codes is presented in a dedicated subsection.
Finally, some results for a protograph-based ARA code are given.

\subsection{Bounds on the code performance}
A lower bound for the CER on the BEC is given by the well-known
Singleton bound, which is matched just by an $(n,k)$ ideal maximum
distance separable (MDS) code:
\begin{equation}\label{eq:singleton}
P_e\geq\sum_{i=n-k+1}^n {n\choose i} \epsilon^i (1-\epsilon)^{n-i}.
\end{equation}
There exist only a few binary codes achieving \eqref{eq:singleton}
with equality. An upper bound on the CER for the random code
ensembles was introduced by Berlekamp \cite{berlekamp:bound}. The
bound can be expressed as:

\begin{equation}\label{eq:random}
\begin{split}
\overline{P}_e\leq\ \sum_{i=0}^{n-k} {n\choose i}
\epsilon^i (1-\epsilon)^{n-i}2^{-(n-k-i)}+\\
    +\sum_{i=n-k+1}^n {n\choose i}
\epsilon^i (1-\epsilon)^{n-i},
\end{split}
\end{equation}
where $\overline{P}_e$ represents the average error probability for
the $(n,k)$ random codes ensemble. Even if \eqref{eq:random}
constitutes an upper bound to the error probability, for
sufficiently-large block lengths such bound can be considered as a
good benchmark for the code performance \cite{MacMullan1998}.

\subsection{Moderate block-size LDPC codes}
\begin{figure}[!p]
\begin{center}
\includegraphics[width=0.9\columnwidth,draft=false]{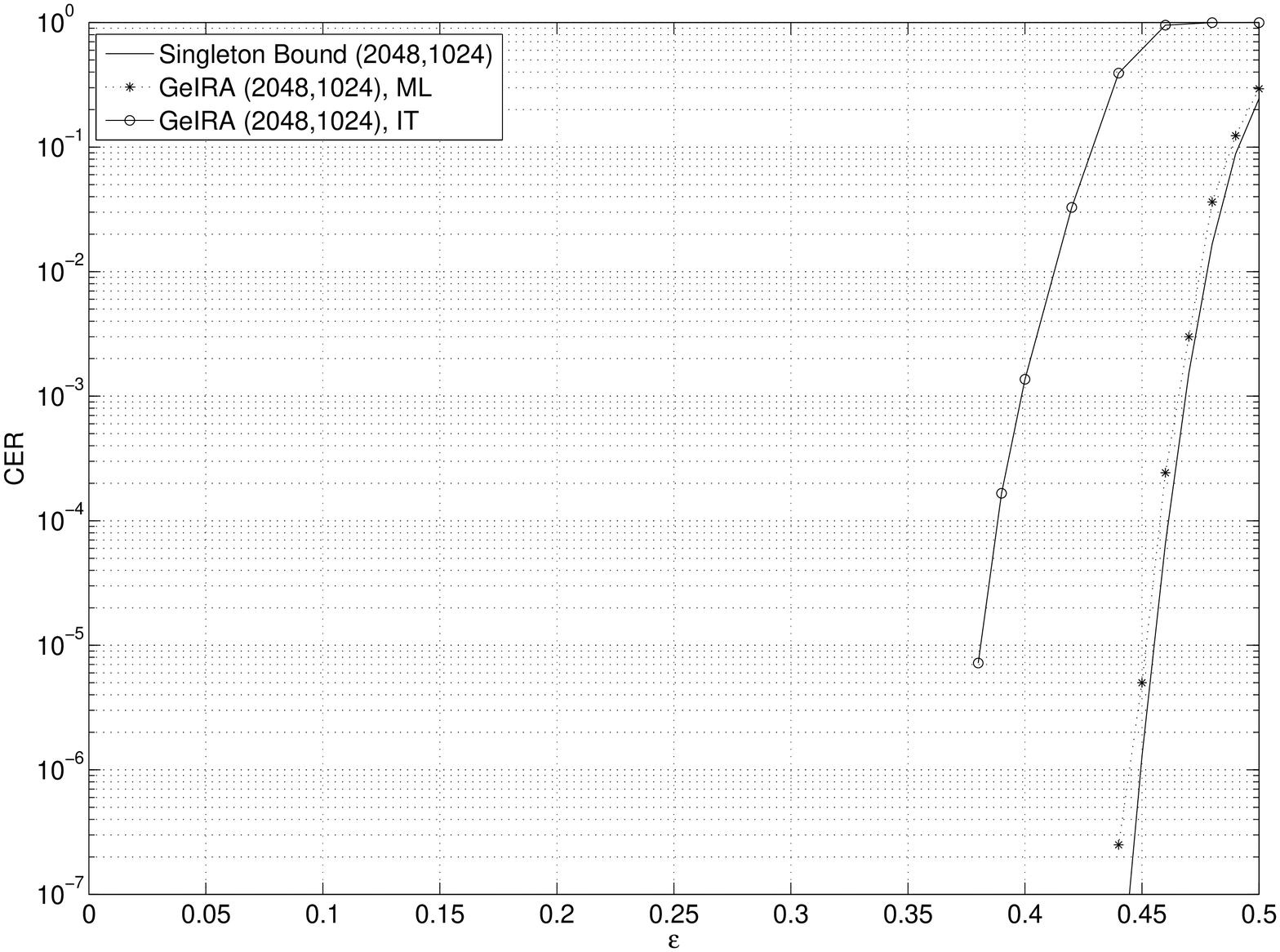}
\end{center}
\caption{Codeword error rate for a (2048,1024) GeIRA code. The solid
line represents the Singleton bound on the CER, while the dotted
line represents the Berlekamp random coding bound.
}\label{fig:Chart_2048_1024}
\end{figure}

\begin{figure}[!p]
\begin{center}
\includegraphics[width=0.9\columnwidth,draft=false]{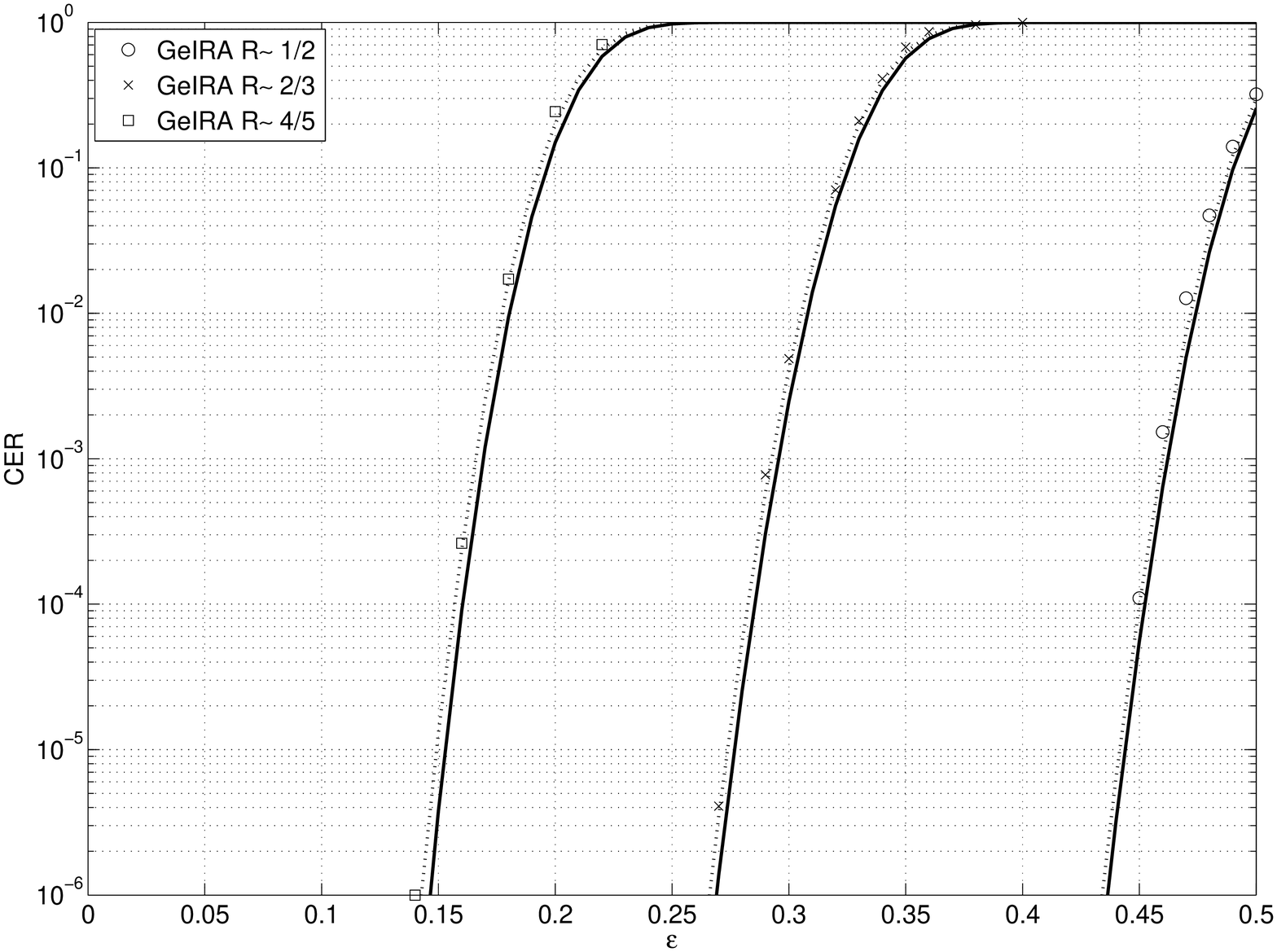}
\end{center}
\caption{Codeword error rates for a family of GeIRA codes with input
block size k=502 and code rates spanning from $1/2$ to $4/5$. The
solid lines represent the respective Singleton bounds on the CER,
while dotted lines represent the respective Berlekamp random coding
bounds.}\label{fig:Chart_family}
\end{figure}

\begin{figure}[!p]
\begin{center}
\includegraphics[width=0.9\columnwidth,draft=false]{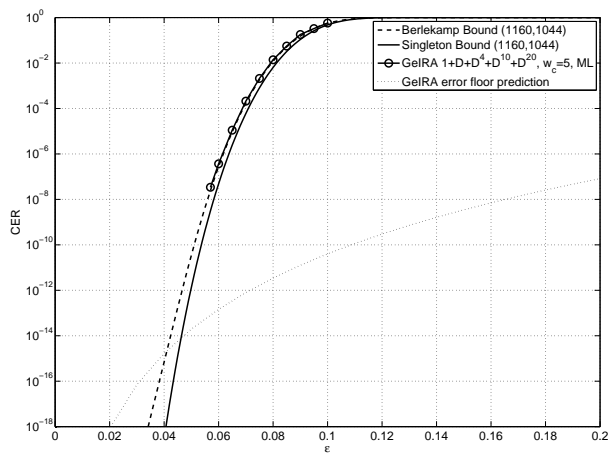}
\end{center}
\caption{Codeword error rate for a (1160,1044) GeIRA code. The
performance is compared the the Berlekamp bound and to the Singleton
bound.}\label{fig:Chart_1160_1044}
\end{figure}

The performance of some moderate-length LDPC codes is provided in
Figures \ref{fig:Chart_2048_1024}, \ref{fig:Chart_family} and
\ref{fig:Chart_1160_1044}. In Figure \ref{fig:Chart_2048_1024}, the
CER for a $(2048,1024)$ GeIRA code from \cite{studio3:MLcommlet} is
presented. The code is picked from an LDPC ensemble with
$\epsilon_{IT}=0.480$ and $\epsilon_{ML}=0.496$. The code
performance, under ML decoding, tightly approaches the Singleton
bound, and pratically matches the Berlekamp bound. The iterative
decoding curve, although not so far from the state-of-the-art for
iteratively-decoded codes, lies quite far from the bound. The
sub-optimality of the IT curve is therefore not due to the code by
itself, but to the sub-optimality of the decoder.

The result is confirmed for a family of rate-compatible GeIRA codes
with code rates ranging from $1/2$ to $4/5$ and input block size
$k=502$ (Figure \ref{fig:Chart_family}. The higher rates are
obtained by puncturing the mother $R=1/2$ code, which has been
derived from the construction proposed in \cite{studio3:MLcommlet}.
For the code rates under investigation, the performance is uniformly
close to the corresponding Singleton bound, down to low codeword
error rates (CER$\simeq10^{-6}$). In Figure
\ref{fig:Chart_1160_1044}, the codeword error rate for a
$(1160,1044)$ $R=9/10$ is shown. The code is a near-regular GeIRA
code with almost constant column weight $w_c=5$ and feedback
polynomial given by $1+D+D^4+D^{10}+D^{20}$. The ML threshold is
$\epsilon_{ML}=0.0994$, while $\epsilon_{IT}=0.0699$. Also in this
case, the error performance curve matches the Berlekamp bound down
to low error rates. The minimum distance of this code (and its
corresponding multiplicity) has been evaluate by
\cite{studio3:mindistLDPC}. An error floor estimation has been
carried out by mean of the truncated union bound on the codeword
error probability, which is given by
\begin{equation}\label{eq:errorfloor}
P_e \simeq A_{min}\epsilon^{d_{min}},
\end{equation}
where $A_{min}$ represents the minimum distance multiplicity. Four
codewords at $d_{min}=11$ have been found, leading to the error
floor estimation provided in Figure \ref{fig:Chart_1160_1044}. Even
if such results represent only an estimation of the actual error
floor, they are quite remarkable. The code performance would in fact
deviate remarkably from the Singleton bound just at error rates
below $10^{-14}$.

\subsection{Comparisons with fixed-rate Raptor codes}
\begin{figure}[!p]
\begin{center}
\includegraphics[width=0.9\columnwidth,draft=false]{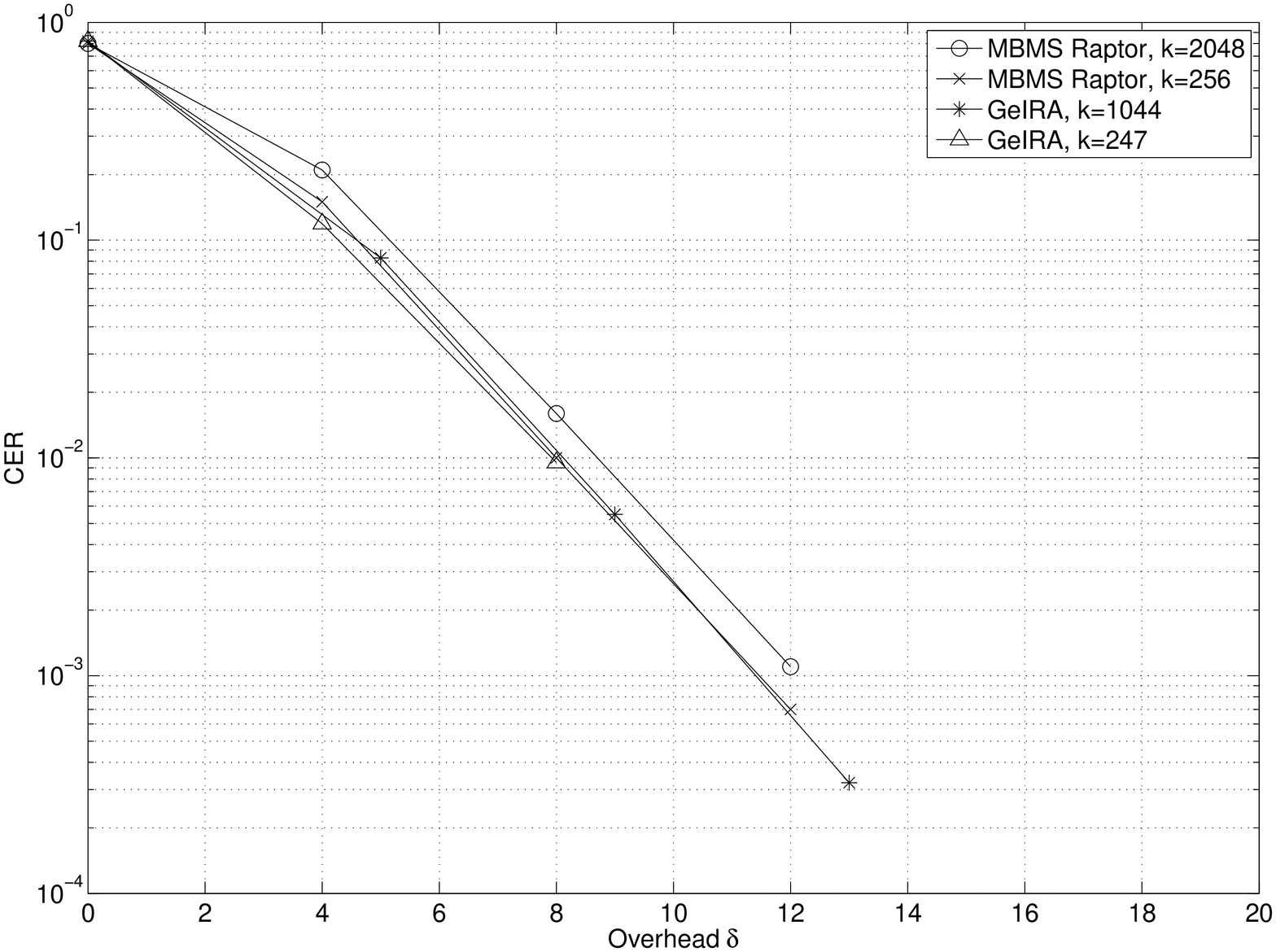}
\end{center}
\caption{Codeword error rate vs. overhead $\delta$ for the MBMS
Raptor code \cite{TransBroad_Luby2007} and for some GeIRA codes,
various input block size.}\label{fig:Chart_Overhead}
\end{figure}

A comparison with fixed-rate Raptor codes specified in the MBMS
standard is provided next. In Figure \ref{fig:Chart_Overhead}, the
decoding failure probability (i.e., the CER) as a function of the
overhead is depicted for the codes specified in \cite{MBMS05:raptor}
and for some GeIRA codes. The overhead $\delta$ is here defined as
the number of codeword symbols that are correctly received in excess
respect to $k$ (recall that $k$ represents the minimum amount of
correctly-received bits allowing successful decoding with an ideal
MDS code). The comparison is carried out for various block sizes.
There is basically no difference in performance between the MBMS
Raptor codes and properly-designed LDPC codes under ML decoding. As
already pointed out in \cite{TransBroad_Luby2007}, the decoding
failure probability vs. overhead does not seem to depend on the
input block size.

\begin{figure}[!p]
\begin{center}
\includegraphics[width=0.9\columnwidth,draft=false]{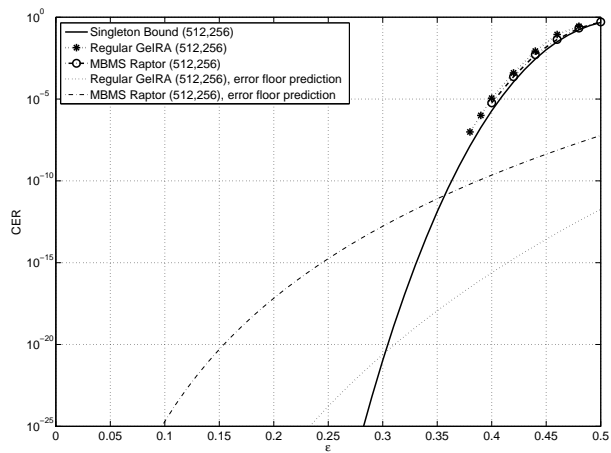}
\end{center}
\caption{Codeword error rates and error floor predictions for
(512,256) Raptor and GeIRA codes.}\label{fig:256_512}
\end{figure}

A comparison between a (512,256) fixed-rate Raptor code and a
near-regular GeIRA code from \cite{studio3:MLcommlet} with constant
column weight $w_c=4$ is provided in Figure \ref{fig:256_512}. In
the waterfall region the two codes exhibit almost the same
performance. A minimum distance estimation according to
\cite{studio3:mindistLDPC} was conducted on the two codes. For the
(512,256) fixed-rate Raptor code, the minimum distance is given by
$d_{min}=25$, with $A_{min}=2$. The lowest Hamming-weight codewords
can be obtained by feeding the encoder with the $k$-bits input
sequences $\mathbf{u}^{(1)}, \mathbf{u}^{(2)}$, where the non-null
bits are $ u^{(1)}_{13}$,
 $u^{(1)}_{21}$,
    $u^{(1)}_{32}$,
    $u^{(1)}_{39}$,
    $u^{(1)}_{63}$,
    $u^{(1)}_{90}$,
    $u^{(1)}_{91}$,
    $u^{(1)}_{95}$,
    $u^{(1)}_{98}$,
   $u^{(1)}_{102}$,
   $u^{(1)}_{115}$,
   $u^{(1)}_{118}$,
   $u^{(1)}_{133}$,
   $u^{(1)}_{142}$,
   $u^{(1)}_{181}$,
   $u^{(1)}_{230}$,
   $u^{(1)}_{243}$,
   $u^{(1)}_{247}$ and
   $u^{(2)}_{6}$,
    $u^{(2)}_{13}$,
    $u^{(2)}_{18}$,
    $u^{(2)}_{75}$,
    $u^{(2)}_{88}$,
   $u^{(2)}_{101}$,
   $u^{(2)}_{123}$,
   $u^{(2)}_{131}$,
   $u^{(2)}_{140}$,
   $u^{(2)}_{143}$,
   $u^{(2)}_{176}$,
   $u^{(2)}_{220}$,
   $u^{(2)}_{231}$,
   $u^{(2)}_{243}$, being $u^{(1)}_{0}$ and $u^{(2)}_{0}$  the first bit
   of $\mathbf{u}^{(1)}$ and $\mathbf{u}^{(2)}$, respectively. For the GeIRA code,
   the estimated minimum distance is $d_{min}=40$, with multiplicity
   $A_{min}=2$. In both cases, the estimated minimum distance is quite large,
   and would permit to achieve very low error floors. For the Raptor
   code, the error floor estimation predicts a deviation from the
   Berlekamp
   bound at CER$\simeq 10^{-11}$, while for the GeIRA code the
   error floor would appear at CER$\simeq 10^{-20}$. The later result is
   quite astonishing, and would suggest the use of the near-regular
   GeIRA construction for applications\footnote{Almost all the current wireless systems adopting erasure correcting codes have requirements which are usually much above the error floor of the Raptor code.} requiring very low error
   floors.
   A final remark on the minimum distance evaluation for fixed-rate
   Raptor codes. The minimum distance evaluation has been applied to
   fixed-rate MBMS Raptor codes with various block lengths. For a
   $(128,64)$ Raptor code, the lowest-weight codeword found by
   \cite{studio3:mindistLDPC} was $14$ ($A_{min}=2$). In the case of
   a $(2048,1024)$ Raptor code, $d_{min}=26$ ($A_{min}=2$).
   Recalling the result for the $(512,256)$ Raptor code
   ($d_{min}=25$), it appears from this preliminary analysis that
   for fixed-rate Raptor codes the minimum distance might scale
   sub-linearly with the block length.

\subsection{ML decoding of a $(1024,512)$ ARA code}
\begin{figure}[!p]
\begin{center}
\includegraphics[width=0.9\columnwidth,draft=false]{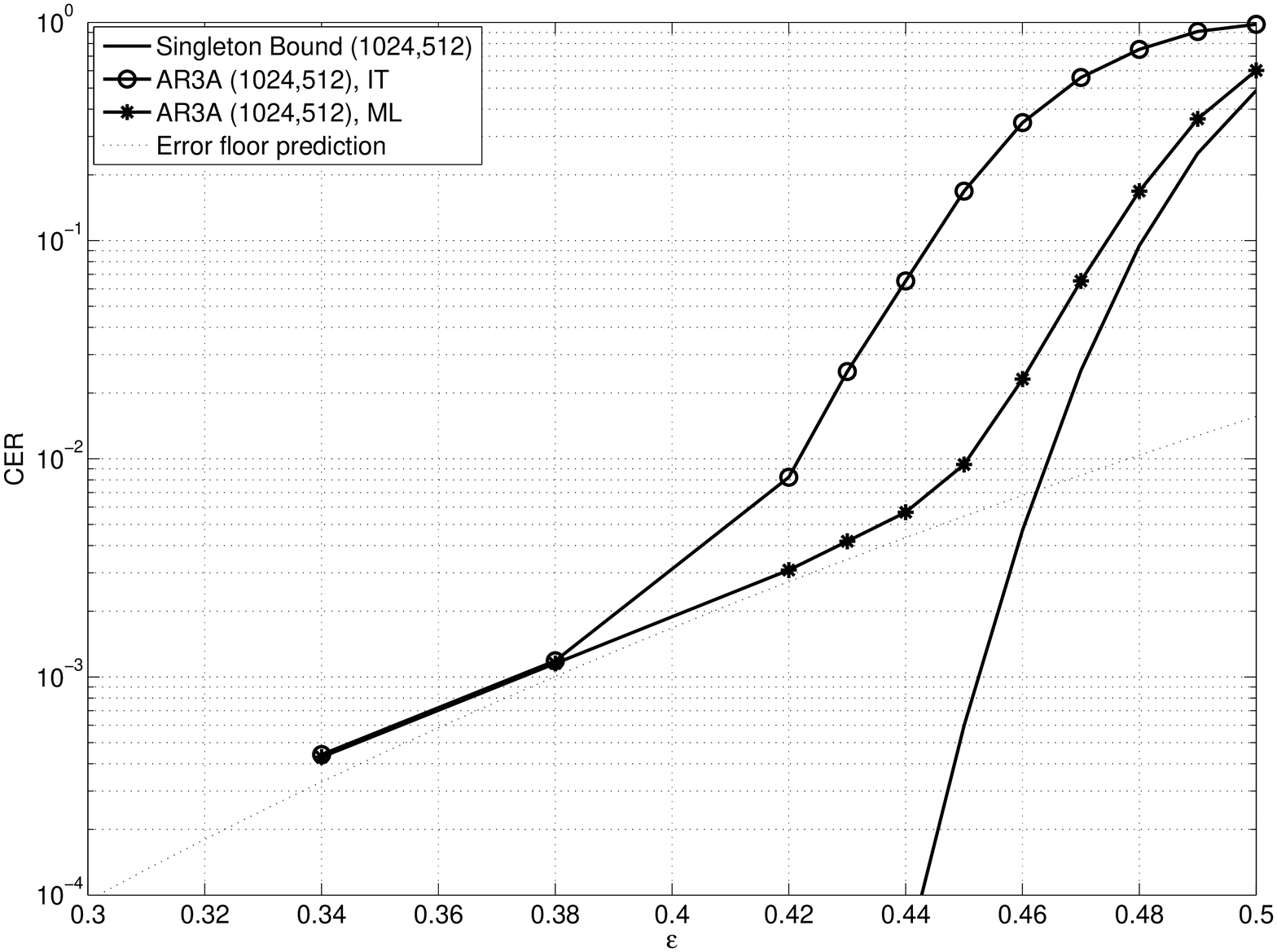}
\end{center}
\caption{Codeword error rates for a (1024,512)
accumulate-repeat-accumulate code under iterative and
maximum-likelihood decoding.}\label{fig:Chart_AR3A}
\end{figure}

In this subsection we provide some numerical results dealing with ML
decoding of a $(1024,512)$ ARA code. The ARA protograph ensemble is
defined by the base matrix \cite{studio3:JCOMSS}
\[
\mathbf{B}=\left(
             \begin{array}{ccccc}
               2 & 1 & 1 & 1 & 0 \\
               1 & 2 & 1 & 1 & 0 \\
               2 & 0 & 0 & 0 & 1 \\
             \end{array}
           \right)
\]
where the first column corresponds to punctured variable nodes. Its
iterative decoding threshold is $\epsilon_{IT}=0.477$. The upper
bound on the ML threshold is $\epsilon_{ML}\leq 0.496$ (see Figure
\ref{fig:GEXIT_ARA}). The code performance is shown in Figure
\ref{fig:Chart_AR3A}, for both iterative and ML decoding. The gain
obtained by the ML decoder in the waterfall region (the error rate
performance is actually quite close to the Singleton bound)
indicates that the bound on the ML threshold is quite tight. Both
the iterative and the ML curves at low error rates present an
evident error floor, due to the presence in the codeword set of $16$
codewords with Hamming weight $10$.

\section{Concluding remarks}\label{sec:CONCLUSIONS}
In this paper we provided some insights on the code design for
ML-decoded LDPC on the erasure channel, together with an overview on
efficient ML decoding algorithms. The complexity on the decoder side
can be kept low with a proper code design. Such approach allows to
design codes with a large flexibility in terms of block lengths and
code rates. A comparison with ML-decoded fixed-rate Raptor codes
(derived from the MBMS specification) has been carried out as well.
The results show that LDPC codes under ML decoding can tightly
approach the bounds down to very low error rates, even for short
block sizes, as their Raptor counterpart. In some cases, the
estimated error floor for the LDPC code is much lower than the
estimated error floor of the corresponding fixed-rate Raptor code.
Since for fixed-rate Raptor codes the error floors are usually very
low, the results achieved with the proposed LDPC are astonishing.
ML-decoded LDPC codes represent therefore a practical tool to
approach the ideal MDS codes performance in many wireless
communications contexts, down to very low error rates, and with
limited decoding complexity.

\section{Acknowledgments}
This research was supported, in part, by the University of Bologna
Grant Internazionalizzazione, and by the EC-IST SatNEx-II project
(IST-27393).

\bibliography{IEEEabrv,asms}

\begin{thebibliography}{10}
\providecommand{\url}[1]{#1}
\csname url@rmstyle\endcsname
\providecommand{\newblock}{\relax}
\providecommand{\bibinfo}[2]{#2}
\providecommand\BIBentrySTDinterwordspacing{\spaceskip=0pt\relax}
\providecommand\BIBentryALTinterwordstretchfactor{4}
\providecommand\BIBentryALTinterwordspacing{\spaceskip=\fontdimen2\font plus
\BIBentryALTinterwordstretchfactor\fontdimen3\font minus
  \fontdimen4\font\relax}
\providecommand\BIBforeignlanguage[2]{{%
\expandafter\ifx\csname l@#1\endcsname\relax
\typeout{** WARNING: IEEEtran.bst: No hyphenation pattern has been}%
\typeout{** loaded for the language `#1'. Using the pattern for}%
\typeout{** the default language instead.}%
\else
\language=\csname l@#1\endcsname
\fi
#2}}

\bibitem{studio3:GallagerBook}
R.~G. Gallager, \emph{Low-Density Parity-Check Codes}.\hskip 1em plus 0.5em
  minus 0.4em\relax Cambridge, MA: M.I.T. Press, 1963.

\bibitem{pfister2005:bounded}
H.~D. Pfister, I.~Sason, and R.~Urbanke, ``Capacity-achieving ensembles for the
  binary erasure channel with bounded complexity,'' \emph{{IEEE} Trans. Inform.
  Theory}, vol.~51, no.~7, pp. 2352--2379, July 2005.

\bibitem{studio3:MLcommlet}
E.~Paolini, G.~Liva, B.~Matuz, and M.~Chiani, ``{Generalized IRA Erasure
  Correcting Codes for Hybrid Iterative / Maximum Likelihood Decoding},''
  \emph{{IEEE} Commun. Lett.}, 2008, accepted for publication.

\bibitem{miller04:bec}
D.~Burshtein and G.~Miller, ``An efficient maximum likelihood decoding of
  {LDPC} codes over the binary erasure channel,'' \emph{{IEEE} Trans. Inform.
  Theory}, vol.~50, no.~11, nov 2004.

\bibitem{DVB-SH}
``{Framing structure, channel coding and modulation for Satellite Services to
  Handheld devices (SH) below 3GHz},'' {Digital Video Broadcasting (DVB)},''
  Blue Book, 2007.

\bibitem{MBMS05:raptor}
{3GPP TS 26.346 V7.4.0}, ``Technical specification group services and system
  aspects; multimedia broadcast/multicast service; protocols and codecs,'' June
  2007.

\bibitem{odlyzko84discrete}
A.~M. Odlyzko, ``Discrete logarithms in finite fields and their cryptographic
  significance,'' in \emph{Theory and Application of Cryptographic Techniques},
  1984, pp. 224--314.

\bibitem{studio3:RiUr01}
T.~Richardson and R.~Urbanke, ``Efficient encoding of low-density parity-ceck
  codes,'' \emph{{IEEE} Trans. Inform. Theory}, vol.~47, pp. 638--656, Feb.
  2001.

\bibitem{Montanari:lifeabovethreshold}
C.~Measson, A.~Montanari, T.~Richardson, and R.~Urbanke, ``Life above
  threshold: From list decoding to area theorem and mse,'' in \emph{Proc. 2004
  IEEE Information Theory Workshop}, San Antonio, USA, October 2004.

\bibitem{Ashikhmin:AreaTheorem}
A.~Ashikhmin, G.~Kramer, and S.~ten Brink, ``Extrinsic information transfer
  functions: Model and erasure channel properties,'' \emph{{IEEE} Trans.
  Inform. Theory}, vol.~50, no.~11, pp. 2657--2673, Nov. 2004.

\bibitem{studio3:proto_LDPC}
J.~Thorpe, ``{Low-Density Parity-Check (LDPC) Codes Constructed Protographs},''
  JPL INP, Tech. Rep. 42-154, Aug. 2003.

\bibitem{studio3:proto_exit2007}
G.~Liva and M.~Chiani, ``{Protograph LDPC codes design based on EXIT
  analysis},'' in \emph{Proc. IEEE Global Communications Conference
  (GLOBECOM)}, Washington, D.C., USA, Nov. 2007.

\bibitem{studio3:ARAGlobecomm}
A.~Abbasfar, K.~Yao, and D.~Disvalar, ``Accumulate repeat accumulate codes,''
  in \emph{Proc. IEEE Globecomm}, Dallas, Texas, Nov. 2004.

\bibitem{liva05:CL}
G.~Liva, E.~Paolini, and M.~Chiani, ``Simple reconfigurable low-density
  parity-check codes,'' \emph{{IEEE} Commun. Lett.}, vol.~9, no.~3, pp.
  258--260, Mar. 2005.

\bibitem{shokrollahi06:raptor}
M.~Shokrollahi, ``Raptor codes,'' \emph{{IEEE} Trans. Inform. Theory}, vol.~52,
  no.~6, pp. 2551--2567, June 2006.

\bibitem{byers02:fountain}
J.~Byers, M.~Luby, and M.~Mitzenmacher, ``A digital fountain approach to
  reliable distribution of bulk data,'' \emph{{IEEE} J. Select. Areas Commun.},
  vol.~20, no.~8, pp. 1528--1540, Oct. 2002.

\bibitem{luby02:LT}
M.~Luby, ``{LT} codes,'' in \emph{Proc. of the 43rd Annual IEEE Symposium on
  Foundations of Computer Science}, Vancouver, Canada, Nov. 2002, pp. 271--282.

\bibitem{Luby2006:Raptor}
M.~Luby, M.~Watson, T.~Gasiba, T.~Stockhammer, and W.~Xu, ``Raptor codes for
  reliable download delivery in wireless broadcast systems,'' in \emph{Proc. of
  2006 {IEEE} Consumer Communications and Networking Conf.}, vol.~1, Jan. 2006,
  pp. 192--197.

\bibitem{berlekamp:bound}
E.~Berlekamp, ``The technology of error-correcting codes,'' \emph{{IEEE}
  Proc.}, vol.~68, pp. 564--593, 1980.

\bibitem{MacMullan1998}
S.~MacMullan and O.M.Collins, ``A comparison of known codes, random codes, and
  the best codes,'' \emph{{IEEE} Trans. Inform. Theory}, vol.~44, Nov. 1998.

\bibitem{studio3:mindistLDPC}
X.-Y. Hu, M.~P.~C. Fossorier, and E.~Eleftheriou, ``On the computation of the
  minimum distance of low-density parity-check codes,'' in \emph{Proc. ICC'04},
  June 2004, pp. 767--771.

\bibitem{TransBroad_Luby2007}
M.~Luby, T.~Gasiba, T.~Stockhammer, and M.~Watson, ``Reliable multimedia
  download delivery in cellular broadcast networks,'' \emph{{IEEE} Transactions
  on Broadcasting}, vol.~53, pp. 235--246, Mar. 2007.

\bibitem{studio3:JCOMSS}
G.~Liva, S.~Song, L.~Lan, Y.~Zhang, W.~Ryan, and S.~Lin, ``{Design of LDPC
  codes: A survey and new results},'' \emph{J. Comm. Software and Systems},
  Sept. 2006.

\end{thebibliography}

\end{document}